\documentclass[12pt,preprint]{emulateapj}

\usepackage{natbib}
\usepackage{lscape}

\shorttitle{Searching for Needles in Haystacks - Looking for GRB $\gamma-$rays with the $Fermi$/LAT Detector}

\slugcomment{Accepted for publication in the Astrophysical Journal}

\begin{document}

\title{Searching for Needles in Haystacks - Looking for GRB $\gamma-$rays with the $Fermi$/LAT Detector}

\author{ C.~W.~Akerlof\altaffilmark{1}, 
\email{cakerlof@umich.edu}
W.~Zheng\altaffilmark{1},
S.~B.~Pandey\altaffilmark{1,2}, 
T.~A.~McKay\altaffilmark{1}
}

\altaffiltext{1} {Randall Laboratory of Physics, Univ. of Michigan, 450 Church Street, Ann Arbor, MI, 48109-1040, USA}
\altaffiltext{2} {Aryabhatta Research Institute of Observational Sciences, Manora Peak, Nainital, India, 263129}

\shortauthors{Akerlof et al. 2010}

\begin{abstract}
Since the launch of the $Fermi~Gamma-ray~Space~Telescope$ on June 11, 2008, 55 gamma-ray bursts (GRBs) have been
observed at coordinates that fall within 66$^\circ$ of the $Fermi$ Large Area Telescope (LAT)
boresight with precise localizations provided by the NASA $Swift$ mission or other satellites. Imposing
selection cuts to exclude low Galactic latitudes and high zenith angles reduces the sample size to 41.
Using matched filter techniques, the $Fermi$/LAT photon data for these fields have been examined for evidence of
bursts that have so far evaded detection at energies above 100 MeV. Following comparisons with similar random background
fields, two events, GRB 080905A and GRB 091208B, stand out as excellent candidates for such an identification. 
After excluding the six bright bursts previously reported by the LAT team, the remaining 35 events exhibit an
excess of LAT ``diffuse'' photons with a statistical significance greater than 2 $\sigma$, independent of
the matched filter analysis. After accounting for the total number of photons in the well-localized fields and including
estimates of detection efficiency, one concludes that somewhere in the range of 11\% to  19\% of all GRBs within the LAT field of view illuminate the detector with
two or more energetic photons. These are the most stringent estimates of the high energy photon content
of GRBs to date. The two new events associated with high energy photon emission have similar ratios of high to low energy fluences as observed previously. This separates them from bursts with similar low energy fluences by a factor of ten, suggesting a distinct class of events rather than a smooth continuum.
\end{abstract}

\keywords{gamma rays: bursts}

\section{Introduction}
One of the more surprising results of the Compton Gamma-Ray Observatory was the EGRET discovery
of an 18 GeV photon associated with GRB 940217 (Hurley {et~al.} 1994). This was not an entirely unique
event; about a half-dozen bursts were seen over the course of the mission with photons above 100
MeV (Catelli et al. 1998, Dingus 2003). Since the GRB spectral energy distribution at lower energies has
been well characterized by a modified power law with peak fluxes at energies of the order of 200
KeV, the existence of photons at energies 10$^4$ times higher puts a significant constraint on any
viable model of the GRB phenomenon (e.g. Band et al. 2009). The obvious question is whether high energy photons are
associated with all GRBs or only with a small sub-class.  We  are now in a unique period when two
space missions are providing GRB coordinates with arc-second accuracy (Swift; Gehrels et al. 2004) and the ability to
measure photon energies to 300 GeV ($Fermi$/LAT; Atwood et al. 2009). Once these two missions cease functioning, the
resources to investigate GRBs over a broad energy range are not likely to be available for decades.
Thus, it is a matter of some urgency to extract the maximum amount of information from the assets
now at hand. Prior to launch of the $Fermi~Gamma-ray~Space~Telescope$, it was possible to speculate
that the LAT instrument would detect more than 200 GRB events per year (Dingus 2003). In the two-year
period since the launch of the $Fermi$ Gamma-ray Space Telescope, the Gamma-ray Burst Monitor
(GBM; Meegan et al. 2009) has reported approximately 475 GRBs, $ie.$ a rate of about 250 per year. Over essentially the
same period, only 17 bursts\footnote{http://fermi.gsfc.nasa.gov/ssc/observations/types/grbs/grb$\_$table/} have been identified by the $Fermi$/LAT. Given that roughly 50\% of all
GBM triggers point to directions within 66$^\circ$ of
the LAT boresight, the fraction of bursts with photons above 100 MeV is apparently not much larger
than 7\%. Thus, it is an important task to apply the best possible techniques to explore whether
this fraction would be significantly bigger if the detection threshold could be lowered. Such
studies might significantly inform the design of future missions while extracting additional
information from rather expensive facilities.

From launch through March 16, 2010, 55 GRBs have been detected (mainly by Swift, as well as INTEGRAL; Winkler et al. 2003,
and AGILE; Giuliani et al. 2008) at celestial coordinates
that were simultaneously viewed by the $Fermi$/LAT. The angular resolutions of the Swift BAT and XRT
are arc-minutes and arc-seconds respectively. Compared with the mean point spread width for the
$Fermi$/LAT of a few degrees, these GRB coordinates are known with zero error. This sample is an
excellent target for statistical methods that can take advantage of the precisely determined source
direction. The basic technique that is employed here is the matched filter, most familiar to those
detecting signals in the time domain. The underlying assumption of this method is that the
characteristics of both the signal and background are $a$ $priori$ known functions of one or more
variables. Since the matched filter maximizes the signal-to-noise ratio, moderate departures from
optimality degrade the filter performance relatively slowly, making this a valuable tool for
investigating the possible existence of faint signals. We present the sample and data processing in
the next section, and details of the signal detection technique in Section 3. Results are given
in Section 4 followed by a brief discussion in Section 5.

\section{Sample Selection and Data Processing}
A selection of 148 precisely located GRBs ($<$1.8') was obtained from the Swift Web
page\footnote{http://swift.gsfc.nasa.gov/docs/swift/archive/grb$\_$table.html/}
devoted to cataloging all bursts. The data correspond to the period from
launch until the end of March 2010. Most of these events were observed by the Swift/XRT, augmented by
a few from INTEGRAL and AGILE. The resulting list was cross-correlated with $Fermi$
spacecraft attitude data obtained from the $Fermi$ Science Support
Center\footnote{http://fermi.gsfc.nasa.gov/ssc/} (FSSC) to identify
only those events within 66$^\circ$ of the LAT boresight.

Having identified the principal focus of this investigation, two sets of LAT photon data
were obtained from the FSSC corresponding to the 55 precisely located GRB fields and to
464 additional fields taken at random on the sky to study the background behavior. Throughout
the following, they will be identified as ``GRB'' or ``random'' data. The FSSC data selection
criteria corresponded to any of the three event classes, ``transient'', ``source'' or ``diffuse'' , photon energies above 100 MeV but below 300 GeV, a zenith angle less than 105$^\circ$ and time intervals when the DATA$\_$QUAL parameter was 1 and the South Atlantic Anomaly parameter, IN$\_$SAA, was 0. The 105$^\circ$ zenith angle limit might be
considered a bit risky but with an orbiting altitude of 565 km, the Earth's limb appears at a
zenith angle of 112.83$^\circ$ which is easily resolvable. For the random
LAT fields, fictitious GRB directions were chosen randomly over the sky but constrained by the less than $66^\circ$
boresight requirement. Further
cuts were imposed of less than $10.5^\circ$ for the angle between the photon and GRB directions and a time
window spanning from $t_0 - 100$ s to $t_0 + 150$ s. With LAT point spread width maxima of the order of
4$^\circ$, the 10.5$^\circ$ cut was imposed to allow up to 2.6 $\sigma$ errors without worrying about the effects
of significant variations in the background flux over the effective solid angle of 0.105 sr. All data fields were
additionally constrained to lie above or below the Galactic plane by more than 10$^\circ$. Such additional criteria
reduced the number of GRB fields from 55 to 41. Six of these have been previously associated with significant
fluxes of energetic photons ($> 100$ MeV) and thus provided guidance about the characteristics of the signals we
were seeking.

\section{Signal Detection Technique}

The central theme of this paper is the use of a matched filter to enhance the signal-to-noise
ratio for the detection of GRB photons in a diffuse background. There are four independent
variables that are involved: energy ($E$), photon angle with respect to the GRB direction ($\theta$),
photon time of arrival ($t$) and photon event class ($c \in \{ 1,2,3\}$). The most obvious consideration is that the photon direction
should be correlated with the GRB coordinates within an angle related to the LAT point spread
function for that specific photon energy. Photons that fall far outside this criterion are
almost certainly background. The energy criterion is less intuitive. The energy spectra for
both the GRB signal and the background are approximately represented by decreasing power laws
with exponents $\Gamma_{GRB}$ and $\Gamma_{back}$ respectively. Unfortunately, the background has a somewhat
harder spectrum with $\Gamma_{back}$ $\simeq$ 1.5 while $\Gamma_{GRB}$ $\simeq$ 2.2. Thus, higher energy photons are more likely
to represent background than signal. Photon directional accuracy is also critical so photons
with small uncertainty (ie. high energy) should be weighted more heavily. The time filter is
the most problematic because GRB light curves lack a uniform shape. Particularly for the most
energetic burst photons, the time delay relative to the low energy trigger can be substantial,
up to thousands of seconds (Hurley {et~al.} 1994). However uniquely identifying such late quanta with
a GRB can be problematic, especially in view of the lack of a distinguishing spectral
characteristic. One feature that is generally observed is a prompt flux within a few seconds
of the low energy trigger. This can be seen in Figure 1 of Ghisellini
{et~al.} (2010). Most of the depicted LAT light curves exhibit a significant flux
within 10 s of the burst. There are a few exceptions such as GRB 090328 so it must be
recognized that not every fish can be caught with the same net. This led us to adopt a
standard GRB light curve described by a broken power-law with a break at $t_b$ = 5 s. Prior to
$t_b$, the flux rises as (t/$t_b$)$^{\alpha_0}$ after which it falls as (t/$t_b$)$^{-\alpha}$. The value for $\alpha_0$ was set
rather arbitrarily at 1/2 while $\alpha$ was picked to be 7/5 in concert with the general behavior of
a large number of GRB afterglows. We emphasize that all the filter parameters were determined
prior to applying them to either the GRB or random data sets.

When we began this analysis, we were unaware of the nuanced distinctions between the three LAT photon
event classes, ``transient'', ``source'' and ``diffuse'', identified as 1, 2 and 3. Since we were looking for a
transient point source, it seemed reasonable to include the ``transient'' quanta on equal footing with the other
two. Only after going through the complete development of the matched filter search did we discover that
the signal content of the ``transient'' photons is extremely small, $\sim 3.5\%$. This was determined by looking at
the event class photon counts for the most intense GRBs, 080916C, 090510, 090902B and 091003. The ratio, $r_{GRB}$, of
``transient'' and ``source'' GRB photons to ``diffuse'' is 0.34 and 0.44 respectively, roughly independent of the particular burst. On the other hand, the same ratios for the selected random LAT fields, $r_{back}$, 
are 9.65 and 0.85. Consequently, we realized that we needed to include a weighting factor, $w_c$, based on event class that
reflected this wide dispersion of information content. Serendiptously, the simple existence of a strong correlation
of GRBs with ``diffuse'' class photons provided an independent verification of the validity of the
matched filter technique.

From these considerations, the mathematical form of the matched filter weight functions is given by:
\begin{equation}
w_E = \frac{1}{4\pi{\sigma}^2(E)}(\frac{E}{E_{th}})^{\Gamma_{back}-\Gamma_{_{GRB}}}
\end{equation}
\begin{equation}
w_\theta = 2 e^{-\frac{{\theta}^2}{2{\sigma}^2(E)}}
\end{equation}
\begin{equation}
w_t =  \Bigg\{ \begin{array}{ll}
       c \cdot (\frac{t}{t_b})^{\alpha_0}  & 0 \leq t \leq t_b \\
       c \cdot (\frac{t}{t_b})^{-\alpha}    & t_b < t \leq t_c \\
       c=2.056382; & t_c = \frac{19}{2}t_b,
    \end{array}
\end{equation}
\begin{equation}
w_c = \begin{array}{ll}
      ({\frac{r_{GRB}(i)}{r_{GRB}(3)}})/({\frac{r_{back}(i)}{r_{back}(3)}}) & 1 \leq i \leq 3
      \end {array}
\end{equation}
\begin{equation}
w(E,\theta,t,c) = w_E \cdot w_\theta \cdot w_t \cdot w_c.
\end{equation} 
The point spread function for the LAT has been specified by the FSSC team in terms of
a half-angle that includes 68\% of a random sample. We have fit this to a power law in
energy (Akerlof \& Yuan 2007) to obtain separate parameter sets corresponding to conversion in either half of
the LAT tracker using the form:
\begin{equation}
{\sigma_{68\%}(E)}=(\frac{E}{E_{th}})^{-\delta}{\sigma}_0+{\sigma}_1,
\end{equation}

For this analysis, we have assumed that the LAT PSF is Gaussian as described above for
$w_{\theta}$ with a characteristic width determined by the containment condition:

\begin{equation}
\label{EQ_PSFSigmaEnergy}
\sigma_{PSF}(E) = \frac{2}{3}{\sigma_{68\%}}(E).
\end{equation}
Although the actual LAT angular error distribution function exhibits non-Gaussian tails, this
approximation is unlikely to affect subsequent conclusions. As indicated above, the
weight value for each photon is computed as a product of the four elements, $w_E$, $w_{\theta}$, $w_t$
and $w_c$. The time filtering algorithm substantially ignores all photons that occur more
than a few tens of seconds after the GRB trigger. For the parameters selected, this
can best be described by an effective T$_{90}$ window time for the signal of 16.864 s
starting 1.624 s after the burst. The sum over all photons, $\sum w_i$, is the statistic
used to identify events with significant fluxes of high energy photons. In assessing
this technique, the mean contribution from background photons turns out to be negligibly
small. The identification problem is driven entirely by infrequent but large fluctuations
of $\sum w_i$ contributed by background quanta. As an additional safeguard, a weight-sharing
measure was introduced to ensure that the entire event weight was not generated by a
single extreme photon. This parameter is defined by:
\begin{equation}
\label{EQ_Coeff_kw}
\zeta = \frac{3(w_{0}\cdot w_{1}\cdot w_{2})^{1/3}}{w_{0} + w_{1} + w_{2}},
\end{equation}
In this formula, $w_0$, $w_1$ and $w_2$ are the maximum three photon weights for a particular
event. The combined statistic, $\zeta \sum w_i$, incorporates both the sum of photon weights and
the additional requirement that no single photon contributes an overwhelming fraction
of the value. The downside is that at least three photons must be associated with an
event, an issue that will arise later.

The significances of the $\sum w_i$ and $\zeta \sum w_i$ statistics were determined by Monte Carlo
simulation of the background. For each field of the GRB data set, 2000 events
were generated using the measured background spectral distribution to determine energy
and uniform distributions over solid angle and time to determine $\theta$ and $t$. The number
of background photons for each field was set by the measured background rate.
By looking at the weight statistics for simulated events with and
without injected GRBs, we gained reasonable confidence that the matched filter algorithm
would correctly identify a significant fraction of real bursts. This work also demonstrated that
the modified sharing-compensated statistic, $\zeta \sum w_i$, was the stronger measure.

As described earlier, 41 GRB fields were selected for examination. These are listed in
Table 2. Six of these events have been previously identified with $>$100 MeV LAT-detected photons.
These are GRB 080916C (Abdo {et~al.} 2009a), GRB 090323
(Ghisellini {et~al.} 2010), GRB 090328A (Ghisellini {et~al.} 2010), GRB 090510 (Abdo {et~al.} 2009b),
GRB 090902B (Abdo {et~al.} 2009c) and GRB 091003 (Ghisellini {et~al.} 2010).

The 35 GRB fields with no previous claims for LAT detections were the target of this investigation. We recognized
that the most convincing argument for a true LAT identification should rely on the statistical distributions
for the matched filter weights in LAT fields with similar characteristics. To increase that number
as much as possible, the LAT fields of view were segmented into 24 circular tiles embedded on a spherical surface
as shown in Figure 1. Each tile subtends a cone with a half-angle of 10.5$^\circ$. This tiling scheme was applied
to both the GRB and random field data sets to realize 697 and 8802 independent directions in space satisfying all the
criteria described previously. Taking advantage of the fact that each field observation was blocked into a 250-s segment,
the number of independent observations was multiplied by five by regarding each 50-s time slice as a separate sample.
Thus, there are 3485 background measurements taken from LAT observations obtained simultaneously with the
candidate GRB fields and an additional 44010 samples taken under similar but not identical conditions. One particular
concern for an analysis of this type is that false positives will selectively occur as the sample photon rate
rises substantially above the mean. Evidence that this is not the case here is shown in Figure 2 which plots the
cumulative distribution of the total number of photons within the LAT FoV over a 250-s interval. These rates
explicitly exclude contamination from photons beyond the 105$^\circ$ zenith angle cut. As shown in the plot,
the two low fluence GRB events reported here are not associated specifically with fields with high ambient
background rates. The
similarity of the distributions for GRB and random fields also shows that the GRB data are not pathological as far as
rates are concerned.

\begin{figure}
  \includegraphics[scale=0.475]{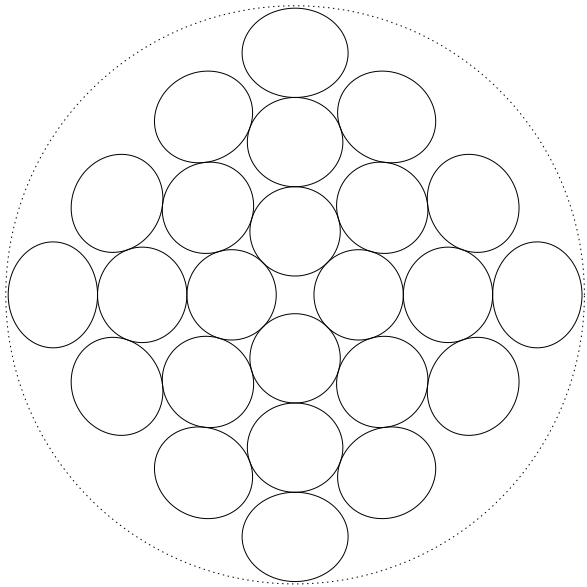}
  \caption{A 10.5$^\circ$ tiling of the LAT field of view. 24 disks are closely confined within the 68$^\circ$
radius of the dotted circle that characterizes the acceptance of the LAT instrument.}
\end{figure}

\begin{figure}
  \includegraphics[scale=0.475]{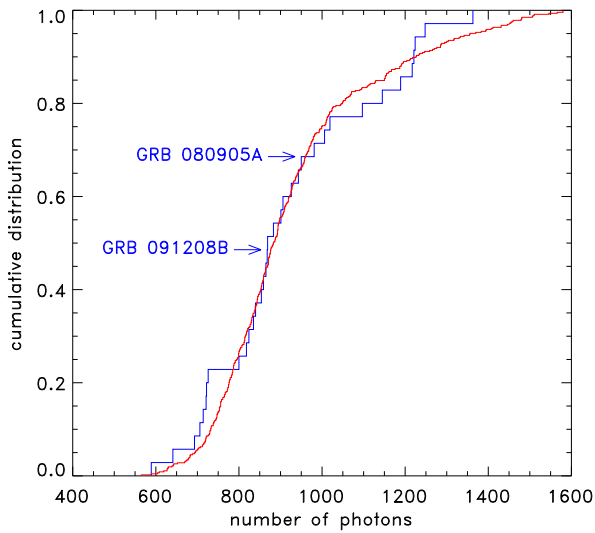}
  \caption{Cumulative distributions of LAT photon rates over 250-s intervals for the GRB (blue) and random
background (red) fields. These rates reflect the entire LAT FoV except for photons that lie outside the
105$^\circ$ zenith angle cut. The rates corresponding to the two claimed GRB detections are indicated by arrows.}
\end{figure}

\section{Results}
The statistical weighting scheme with the additional fractional sharing requirement
clearly identified four of the six events previously associated with high energy gamma-ray
emission. The exceptions were GRB 090323 and GRB 0903028A for which photon counts were very sparse
during the first 50 seconds following the burst trigger. Of the two, 090323 was marginally detected by
the matched filter technique but GRB 0903028A provided no sign of a signal whatsoever. This gives some order of magnitude
indication of the likely efficiency of the method.
Three of the six were extremely intense with
100 or more energetic photons (GRB 080916C, GRB 090510 and GRB 090902B). As will be discussed later, the
existence of such a substantial bright fraction is a rather surprising situation.

Most significantly, our analysis identified two additional bursts, GRB 080905A and GRB 091208B,
as probably associated with energetic photon detections in the LAT. The best estimate for the
probability of these occuring by chance alone was obtained by performing identical searches on
random LAT fields with the same criteria. Thus, 1 out of 44010 random fields generated matched filter weights
exceeding the value for GRB 080905A  and 11 similar fields generated weights exceeding the value for GRB 091208B.
Multiplying by a trials factor of 35 for the number of localized fields considered, the associated probabilities
are $8.0 \times 10^{-4}$ and $8.7 \times 10^{-3}$. To check that these correlations were simply not due
to preferentially higher background rates for the GRB exposures, we also performed similar calculations for
the LAT data confined to $\sim$ 20 uncorrelated directions and five independent time intervals from the same GRB data sets. All of these results are plotted as cumulative distributions in Figure 3. The statistical similarity of the
GRB off-axis and random field data demonstrates that the GRB data set is not correlated with anomalous environmental
conditions such as higher cosmic ray background rates. A list of photons associated with these two
events is provided in Table 1.

Two independent observations at significance levels of $0.1\%$ and $1\%$ already make this a reasonably credible
result. However, as we discovered the importance of LAT "diffuse" photons for identifying point sources such as GRBs,
we realized that independent of the matched filter weight analysis, a simple count of the number of such quanta
was statistically quite significant. Thus for the 35 well-localized GRB fields that don't include the previously
reported LAT-detected GRBs, the total number of ``diffuse'' photons is 25 for a rate of $0.714 \pm 0.143$.
For the GRB off-axis fields and the random fields, the comparable numbers are 1384 for 3485 fields and 17414 for 44010
to yield per field rates of $0.397 \pm 0.011$ and $0.396 \pm 0.003$. The random probability of such differences
is less than $2.7 \%$. As can be seen from the lists of photon counts in Tables 1 and 2, the two new GRB identifications
are strongly correlated with
these ``diffuse'' photons as well. Thus, the combined  probablility that these claims might be randomly
generated must be less than $~1 \times 10^{-5}$.

GRB 080905A is a short burst with optical counterparts and
a redshift of 0.128 (Rowlinson et al. 2010) while GRB 091208B is a long burst also with
optical counterparts (Pagani et al. 2010) and a redshift of 1.063 (Wiersema
et al. 2010; Perley et al. 2010). As listed in Table 1, the estimated number of $>$100 MeV photons
for each event is about 3. This is substantially lower than any previous LAT GRB identifications.
Estimating the corresponding fluence of $>$100 MeV photons raised some interesting statistical issues
that are perhaps not well appreciated. To first order, one would anticipate simply adding up all the
energies of the detected quanta and dividing by the effective detector area. Although this produces a number,
it bears only a casual relation to the intrinsic properties of the GRB in question. The reason is that the
spectral number distribution, $dN/dE$ follows a power law with exponent between -3 and -2. Thus the mathematical mean is defined but the variance is infinite. This leads to the conclusion that sums of energies drawn from the identical
parent population will vary by factors of the order of unity, independent of the number of photons in the
sample. In our case, the statistical uncertainty is already substantial since only 3 photons are detected for
each event. Thus, we opted for a solution that invoked the least number of new assumptions: we adopted the LAT
fluence values for bright bursts computed by Ghisellini et al. (2010) and listed in their Table 1.
We next assumed that the spectral distribution for all bursts was an approximately universal function of energy so that fluences would scale with photon number. This led to estimates of the LAT fluence value of $3 \times 10^{-6}~ergs/cm^2$ for both events. Unfortunately, the proportionality of fluence to photon counts is not very evident. As can be seen in Table 2
for the two bursts, GRB 090510 and GRB 090902B, for essentially the same number of photons, the fluences differ
by a factor of 16.
This problem may require a better realization of how fluences should be estimated. For future purposes when dealing
with events with more numerous quanta, there is a robust measure of fluence: multiply
the number of photons by the median energy. In the continuum limit, a power-law distribution generates the following relation
between $E_{mean}$ and $E_{median}$ over the interval from threshold to infinity:
\begin{equation}
{E_{mean}}=\frac{\Gamma - 1}{\Gamma - 2}\frac{1}{2^{\Gamma - 1}}E_{median}
\end{equation}
Thus a robust proxy for gamma-ray burst fluence can be obtained by multiplying this estimate of $E_{mean}$ by the
number of detected photons.

\begin{figure}
  \includegraphics[scale=0.475]{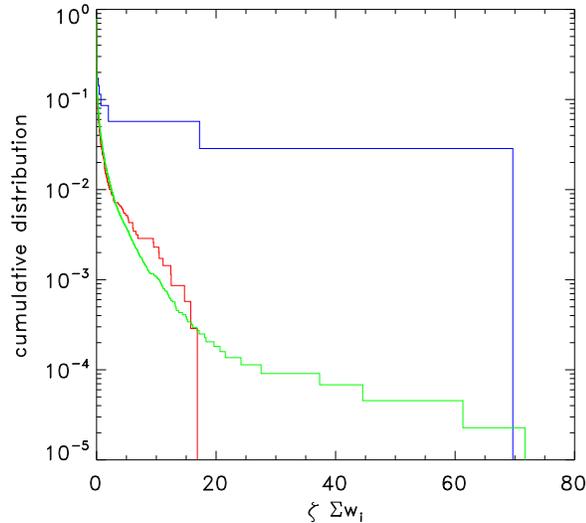}
  \caption{Complements of the cumulative distributions for ${\zeta}{\sum}w_i$
           for 35 well-localized GRB fields (blue), 3485 random fields obtained nearly simultaneously with
           the GRB data (red) and 44010 random fields obtained at random times (green).}
\end{figure}

\begin{deluxetable}{crrrrr}
 \tabcolsep 0.4mm
 \tablewidth{0pt}
 \tablecaption{~ List of high energy photons for GRB 080905A \& GRB 091208B}\label{tab:Tab_Highweight}
  \tablehead{\colhead{} & \colhead{} & \colhead{} & \colhead{} & \colhead{} & \colhead{}}
\\
\multicolumn{6}{l}{GRB 080905A~~~~~${\zeta}$ = 0.2118~~~~~${\zeta}{\sum}w_i$ = 69.66}\\
\\
 $i$ & $t$ (s) & $\theta$ ($^\circ$) & $E$ (MeV) & $c$ & $w_i$ \\
\\
  1 & 10.019 &  0.21 &  474 &  3 &  267.07   \\
  2 &  5.661 &  2.27 &  181 &  3 &   61.06   \\
  3 &  6.727 & 10.41 &  103 &  2 &    0.77   \\
\\
\multicolumn{6}{l}{GRB 091208B~~~~~${\zeta}$ = 0.8442~~~~~${\zeta}{\sum}w_i$ = 17.23}\\
\\
 $i$ & $t$ (s) & $\theta$ ($^\circ$) & $E$ (MeV) & $c$ & $w_i$ \\
\\
  1 & 17.932 &  2.06 &  131 & 3 &  11.18   \\
  2 & 20.985 &  2.72 &  125 & 2 &   3.94   \\
  3 &  4.968 &  1.66 &  544 & 1 &   3.02   \\
  4 &  6.364 &  1.16 & 1179 & 1 &   2.07   \\
  5 & 19.735 &  7.17 &  197 & 3 &   0.20   \\
\\
\tableline
\end{deluxetable}

The fact that $>$100 MeV photons have been observed with both long and short bursts
suggests that the production mechanism is generic to relativistic jets, independent of
the detailed structure of the progenitor system. So far, most of the high energy photon
observations have been associated with large fluences at the lower energies
measured by the Swift/BAT and the $Fermi$/GBM detectors. To explore such fluence correlations,
Figure 4 shows a scatterplot of LAT fluences vs. GBM fluences for the six bright bursts
previously identified by the LAT collaboration, the two bursts uncovered by the
statistical analysis described in this paper and the co-added sum of nine bursts with no
strong perceptible radiation above 100 MeV. For these purposes, GBM fluences were defined by
photons in the range of 8 to 1000 KeV within the $T_{90}$ time interval. The LAT fluences
were obtained from Ghiselini et al. (2010) and the scaling method described previously.

The salient feature of Figure 4 is the overall linear correlation of the LAT and
GBM fluences with a ratio of unity between the two. The short burst identified by our analysis,
GRB 080905A, lies above the trend line, a feature previously noted for this class of events
(see discussion of GRB 081024B in Abdo {et~al.} 2010 and GRB 090510 in Ackermann {et~al.} 2010). GRB 091208B
is completely consistent with the higher fluence events. An interesting aspect of this graph is
the estimate for the total co-added LAT fluence for nine bursts with GBM fluences roughly equal or greater to GRB 091208B
but which lie a factor of ten lower but with rather large errors. If one believes that all GRBs have a similar generic behavior, it is possible that a large fraction of these events have delayed high energy emission that evades our estimates.
It is more likely that there is a real population difference between events like 080916C, 090926A and 091208B
and the majority of all other GRBs. Examining this question with larger data sets may shed some light on the
mechanisms that drive high energy emission processes.

Although the statistics are admittedly crude, we can use this analysis to set some limits on the fraction of high energy
photon GRBs at the present levels of sensitivity. Starting with a sample of 41 well-localized events, we
have identiifed two with high statistical significance. The remaining number of LAT ``source'' and ``diffuse'' photons
above background is no more than
approximately three. The efficiency of this technique is about $50\%$ for a total so as many as
16 GRBs out of 41 fields could be associated with $>$ 100 MeV photons. Thus, instead of 17 LAT-detected GRBs over a two-year period, there might be 45 or $19\%$ of all viewable GBM triggers with a lower limit of
25 or $11\%$. That is lower than any estimates prior to the $\it Fermi$ launch.

\begin{figure}
  \includegraphics[scale=0.475]{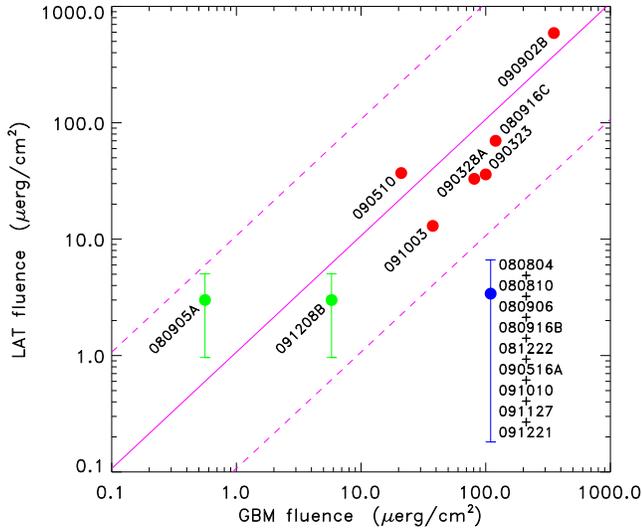}
  \caption{Plot of high energy (LAT) fluences vs. low energy (GBM) fluences for the bursts considered in this
paper. The six red points indicate the events previously reported by the LAT team. The two green points were
identified by the matched filter technique. The solid magenta line is a linear fit to the six LAT-detected
bursts, 080916C, 090323, 090328A, 090510, 090902B and 091003. The dashed lines above and below indicate LAT
fluences ten times greater and smaller.The blue point is the co-added sum of nine GRBs with GBM
fluences comparable or greater to GRB 091208B, ie. greater than 5 $\mu$ergs/cm$^2$. The abscissa is the sum of
all 9 GBM fluences and the ordinate is estimated from the total of all event class 2 \& 3 photons in the corresponding
burst fields. This indicates a significant population with high energy fluences below the
approximately 1:1 fluence ratio represented by the diagonal solid magenta line.
}
\end{figure}

\section{Discussion}
By using optimal signal detection techniques, we have demonstrated that the existence of GRB high energy photon
emission can be extended downwards in brightness by a significant factor. Increasing the dynamic range
over which such phenomena can be measured puts a number of constraints on physical models that can describe these
events. As we have shown, these statistical methods can set interesting limits on the behavior of ensembles of events
at lower levels than is possible for single isolated bursts, raising serious questions about the homogeneity of
the GRB phenomenon. After another two years in orbit, the $Fermi$ data will increase enough to allow more careful
re-examination of the initial conclusions presented here.

The most surprising aspect of this work is the very small number of GRBs that could be positively identified with
high energy emission as a result of the substantially lower fluence thresholds. Almost any model would anticipate an
increase in number proportional to the inverse of the ratio of detection thresholds raised to the power of unity or greater.
Monte Carlo estimates indicate that our detection efficiency was of the order of 50\% for the two events reported.
This does not come close to explaining the apparent deficit of low-intensity events compared to the number of bursts similar to
GRBs 080916C, 090510 and 090902B . We look forward to pursuing this problem during this precious and unique period when both
$Fermi$ and $Swift$ are working magnificently.

\vspace{0.1cm}

\acknowledgments
 
We thank Chris Shrader, director of the $Fermi$ Science Support Center for his considerable help in obtaining and interpreting the $Fermi$ mission data products. Fang Yuan provided valuable assistance in the initial process of learning
how to access and manipulate the $Fermi$ data. This research is supported by NASA grant NNX08AV63G and NSF
grant PHY-0801007.

\clearpage
\LongTables
\begin{landscape}

\begin{deluxetable}{rlrccrrrrrrrrrrrr}
 \tabcolsep 0.0mm
 \tablewidth{10pt}
 \tablecaption{~ List of 41 precisely located LAT-observed GRBs}\label{tab:Tab_OurSample}
\tablehead{\colhead{} & \colhead{GRB} & \colhead{$T_{90}$} & \colhead{UT}  & \colhead{Trigger$^a$} & \colhead{$\alpha^b$} &
 \colhead{$\delta^c$} & \colhead{$\theta_{zenith}^d$} & \colhead{$\theta_{bore}^e$} & \colhead{$b^f$} &  \colhead{$n_{total}^g$} &
 \colhead{$n_1^h$}  & \colhead{$n_2^h$}  & \colhead{$n_3^h$}  & \colhead{${\zeta}{\sum}w_i$$^i$} &  \colhead{$S_{GBM}$$^j$} &
 \colhead{$S_{LAT}$$^k$} \\
 \colhead{} & \colhead{} & \colhead{(s)} & \colhead{} & \colhead{} & \colhead{($^\circ$)} & \colhead{($^\circ$)} &
 \colhead{($^\circ$)} & \colhead{($^\circ$)} & \colhead{($^\circ$)} & \colhead{} & \colhead{} & \colhead{} & \colhead{}  &
 \colhead{} & \colhead{($\mu erg/cm^2)$} & \colhead{($\mu erg/cm^2)$}
}                                                                                                                
\\
\multicolumn{17}{c}{Previously reported GRBs with LAT$-$detected photons}\\
\\                                                    
  1 & 080916C &  100.9     & 00:12:45  &  $Fermi$/GBM & 119.847 & -56.638 & 83.657 & 48.838 & -13.760 &  916 & 26 & 31 &  85 &  7910.540 & 120.00    & 70  \\
  2 & 090323  &  70.0      & 00:02:42  &  $Fermi$/GBM & 190.709 &  17.054 & 69.558 & 59.538 &  79.731 &  763 &  2 &  0 &   2 &    10.570 & 100.00    & 36  \\
  3 & 090328A &  100.0     & 09:36:46  &  $Fermi$/GBM &  90.665 & -41.883 & 71.005 & 64.538 & -26.384 &  932 &  4 &  2 &   1 &     0.606 & 80.90     & 33  \\
  4 & 090510  &  0.3       & 00:23:00  &  351588      & 333.553 & -26.583 & 52.541 & 13.468 & -55.074 &  898 & 26 & 48 & 112 & 31389.817 & 21.00     & 37  \\
  5 & 090902B &  21.0      & 11:05:08  &  $Fermi$/GBM & 264.939 &  27.325 & 41.225 & 50.954 &  26.911 & 1088 & 54 & 49 &  93 &  8530.459 & 352.00    & 590  \\
  6 & 091003  &  21.1      & 04:35:45  &  $Fermi$/GBM & 251.520 &  36.625 & 62.298 & 12.306 &  40.055 & 1062 &  8 &  2 &   7 &  1107.330 & 37.60     & 13  \\
\\                                          
\hline                                         
\\
\multicolumn{17}{c}{New candidate GRBs with LAT$-$detected photons}\\
\\
  7 & 080905A &  1.0       & 11:58:54  &  323870      & 287.674 & -18.880 & 30.418 & 29.860 & -12.605 &  950 &  2 &  1 &   3 &    69.660 & 0.56      & 3   \\
  8 & 091208B &  14.9      & 09:49:57  &  378559      &  29.392 &  16.890 & 28.815 & 55.885 & -43.150 &  868 &  2 &  1 &   3 &    17.227 & 5.80      & 3   \\
\\                                    
\hline                                       
\\                                             
\multicolumn{17}{c}{GRBs with no significant LAT$-$detected photons} \\
\\
  9 & 080804  &  34        & 23:20:14  &  319016      & 328.668 & -53.185 & 83.852 & 58.149 & -48.369 &  641 &  0 &  0 &   0 &         0 & $^*$6.00  &        \\
 10 & 080810  &  106       & 13:10:12  &  319584      & 356.794 &  -0.320 & 29.913 & 62.123 & -59.066 & 1019 &  1 &  0 &   0 &         0 & 6.90      &        \\
 11 & 080906  &  147       & 13:33:16  &  323984      & 228.042 & -80.518 & 71.429 & 36.599 & -19.264 & 1189 & 11 &  1 &   1 &     0.034 & $^*$5.83  &        \\
 12 & 080916B &  32        & 14:44:47  &  324907      & 163.665 &  69.065 & 55.357 & 27.941 &  44.707 & 1248 &  9 &  0 &   1 &     1.975 & 10.90     &        \\
 13 & 080928  &  280       & 15:01:32  &  326115      &  95.070 & -55.200 & 76.016 & 45.239 & -26.313 &  726 &  3 &  0 &   1 &         0 & 3.50      &        \\
 14 & 081003A &  $\sim$30  & 13:46:12  &  INTEGRAL    & 262.391 &  16.571 & 21.858 & 56.387 &  25.363 &  926 &  2 &  0 &   1 &         0 & -         &        \\
 15 & 081012  &  29        & 13:10:23  &  331475      &  30.201 & -17.638 & 68.915 & 61.566 & -71.399 &  721 &  1 &  0 &   0 &         0 & 4.30      &        \\
 16 & 081016B &  2.6       & 19:47:14  &  331856      &  14.564 & -43.530 & 83.851 & 65.064 & -73.540 & 1224 &  2 &  0 &   0 &         0 & $^*$0.17  &        \\
 17 & 081029  &  270       & 01:43:56  &  332931      & 346.773 & -68.156 & 79.789 & 57.063 & -46.109 &  981 &  3 &  0 &   1 &     0.001 & $^*$3.50  &        \\
 18 & 081104  &  59.1      & 09:34:42  &  333666      & 100.489 & -54.720 & 66.142 & 32.627 & -23.173 &  835 &  3 &  0 &   1 &         0 & $^*$3.33  &        \\
 19 & 081118  &  67        & 14:56:36  &  334877      &  82.593 & -43.301 & 65.099 & 30.522 & -32.485 &  706 &  2 &  1 &   0 &     0.001 & 0.11      &        \\
 20 & 081127  &  37        & 07:05:08  &  335715      & 332.064 &   6.851 & 20.654 & 52.608 & -37.920 &  869 &  2 &  0 &   0 &         0 & $^*$0.82  &        \\
 21 & 081222  &  24        & 04:53:59  &  337914      &  22.741 & -34.096 & 29.145 & 51.259 & -79.017 &  714 &  7 &  1 &   2 &     0.447 & 13.50     &        \\
 22 & 090113  &  9.1       & 18:40:39  &  339852      &  32.057 &  33.429 & 21.263 & 32.518 & -26.760 &  883 &  3 &  0 &   2 &         0 & $^*$1.27  &        \\
 23 & 090407  &  310       & 10:28:25  &  348650      &  68.979 & -12.679 & 13.189 & 34.739 & -35.741 &  800 &  3 &  1 &   0 &         0 & $^*$1.83  &        \\
 24 & 090516A &  210       & 08:27:50  &  352190      & 138.261 & -11.854 & 11.567 & 31.509 &  24.261 & 1217 &  6 &  0 &   0 &         0 & 23.00     &        \\
 25 & 090518  &  6.9       & 01:54:44  &  352420      & 119.954 &  -0.759 & 33.312 & 49.588 &  14.806 &  854 &  2 &  0 &   1 &         0 & 1.60      &        \\
 26 & 090519  &  64        & 21:08:56  &  352648      & 142.279 &  -0.180 & 51.625 & 50.424 &  34.312 &  693 &  3 &  1 &   0 &     0.018 & $^*$2.00  &        \\
 27 & 090529  &  $>$100    & 14:12:35  &  353540      & 212.469 &  24.459 & 80.559 & 23.511 &  72.163 &  589 & 11 &  0 &   0 &     0.017 & $^*$1.13  &        \\
 28 & 090702  &  $\sim$10  & 10:40:37  &  INTEGRAL    & 175.897 &  11.502 & 80.386 & 62.327 &  67.683 &  859 &  1 &  0 &   0 &         0 & -         &        \\
 29 & 090708  &  15.0      & 03:38:15  &  356776      & 154.632 &  26.616 & 79.492 & 56.487 &  56.031 & 1006 &  1 &  0 &   0 &         0 & 0.40      &        \\
 30 & 090709B &  27.2      & 15:07:42  &  356912      &  93.522 &  64.081 & 85.836 & 46.897 &  20.304 &  818 &  2 &  1 &   1 &     0.025 & 1.30      &        \\
 31 & 090712  &  145       & 03:51:05  &  357072      &  70.097 &  22.525 & 60.746 & 33.281 & -15.676 &  865 &  8 &  1 &   0 &     0.715 & 4.20      &        \\
 32 & 090728  &  59        & 14:45:45  &  358574      &  29.653 &  41.633 & 90.838 & 60.660 & -19.505 & 1221 &  2 &  0 &   0 &         0 & $^*$1.67  &        \\
 33 & 090813  &  7.1       & 04:10:43  &  359884      & 225.779 &  88.568 & 72.707 & 35.418 &  28.328 &  824 &  2 &  0 &   0 &         0 & 3.50      &        \\
 34 & 091010  &  8.1       & 02:43:09  &  AGILE       & 298.666 & -22.518 & 68.465 & 55.714 & -23.488 & 1145 &  3 &  0 &   3 &     0.001 & 10.90     &        \\
 35 & 091127  &  7.1       & 23:25:45  &  377179      &  36.583 & -18.953 & 54.011 & 25.423 & -66.738 &  840 &  8 &  0 &   1 &     0.130 & 18.70     &        \\
 36 & 091202  &  $\sim$50  & 23:10:12  &  INTEGRAL    & 138.834 &  62.550 & 69.166 & 23.256 &  40.214 &  906 &  7 &  0 &   1 &     0.289 & -         &        \\
 37 & 091221  &  68.5      & 20:52:52  &  380311      &  55.797 &  23.241 & 87.598 & 53.620 & -24.755 &  943 &  4 &  0 &   0 &         0 & 13.80     &        \\
 38 & 100111A &  12.9      & 04:12:49  &  382399      & 247.048 &  15.551 & 48.142 & 33.854 &  38.604 & 1363 &  6 &  0 &   2 &         0 & 1.50      &        \\
 39 & 100206A &  0.12      & 13:30:05  &  411412      &  47.162 &  13.157 & 49.423 & 44.608 & -37.742 & 1097 &  6 &  0 &   0 &     0.001 & 0.93      &        \\
 40 & 100212A &  136       & 14:07:22  &  412081      & 356.418 &  49.494 & 56.829 & 23.422 & -11.989 &  901 &  4 &  0 &   0 &         0 & 0.38      &        \\
 41 & 100316D &  $\sim$740 & 12:44:50  &  416135      & 107.628 & -56.255 & 90.330 & 50.079 & -19.759 &  722 &  5 &  0 &   0 &         0 & $^*$0.50  &        \\
\hline
\tablenotetext{a}{Swift trigger number if numeric; otherwise satellite name.}
\tablenotetext{b}{GRB Right Ascension.}
\tablenotetext{c}{GRB Declination.}
\tablenotetext{d}{Angle between local orbit zenith direction and GRB.}
\tablenotetext{e}{Boresight angle of GRB with respect to the LAT principal axis.}
\tablenotetext{f}{GRB Galactic latitude.}
\tablenotetext{g}{Total number of LAT-detected photons over 250-s interval excluding those beyond $105^\circ$ zenith cut.}
\tablenotetext{h}{$n_1$, $n_2$ and $n_3$ are the number of photons within $10.5^\circ$ of GRB direction in 47.5-s window.}
\tablenotetext{i}{Total event matched filter weight including the sharing parameter, $\zeta$.}
\tablenotetext{j}{GBM fluence, $S_{GBM}$, calculated as described in the main text. $^*$ indicates an entry
with no available GBM fluence measurement; the value is estimated from the BAT fluence by multiplying by a factor of 1.67
as derived from a joint sample of BAT and GBM measurements. Most other fluence values were taken from Table 2 in
Guetta et al. 2010. The exceptions were obtained from the following references:
GRB 080905A, Bissaldi et al., 2008;
GRB 081118, Bhat et al., 2008;
GRB 090510, Guiriec et al., 2009;
GRB 090518, Paciesas, 2009;
GRB 100206A, von Kienlin 2010.}
\tablenotetext{k}{LAT fluence, $S_{LAT}$, calculated as described in the main text.}
\end{deluxetable}
\clearpage
\end{landscape}
\end{document}